# Group Movie Selection using Multi-channel Emotion Recognition

Elnara Kadyrgali, Adilet Yerkin, Yerdauit Torekhan, Pakizar Shamoi*
School of Information Technology and Engineering
Kazakh-British Technical University
Almaty, Kazakhstan
*p.shamoi@kbtu.kz

*Abstract*—Social activities often done in groups include watching television or movies. Choosing a film that appeals to the emotional inclinations of a varied group can be tricky. One of the most difficult aspects of making group movie suggestions is achieving agreement among members. At the same time, emotion is the most important component that connects the film and the viewer. Current research proposes a methodology for group movie selection that employs emotional analysis from numerous sources, such as film posters, soundtracks, and text. Our research stands at the intersection of emotion recognition technology in music, text, color images, and group decision-making, providing a practical tool for navigating the complex dynamics of film selection in a group setting. The survey participants were given emotion categories and asked to select the emotions that best suited a particular movie. Preliminary comparison results between real and predicted scores show the effectiveness of using emotion detection for group movie recommendation. Such systems have the potential to enhance movie recommendation systems.

## I. Introduction

In film selection, achieving consensus among individuals with diverse preferences can be challenging.

Emotion is the most vital factor that connects the movie and the human. In contemporary entertainment, viewing a movie is more than just visual and auditory stimulation; it explores the complex domain of human emotions.

Watching television or movies are examples of social activities typically done in groups [3]. The difficulty of selecting a film that satisfies a diverse group's overall emotional preferences can pose a significant challenge. Facilitating consensus among group members is one of the most significant challenges in group movie recommendations. Emotions are crucial in this perspective, as we primarily watch movies to experience emotions.

Emotion recognition studies can be classified based on approaches used in research. Studies of the first category focus on analyzing a single component to detect the emotion. In contrast, the second ones use multi-componential approaches in emotion analysis, integrating more human-like applications.

Several studies have examined the problem of group movie recommendation using various methods.

The study [1] introduces "Happy movie," a Facebook-integrated application that enhances group movie recommendations by using personality, social trust, and past preferences, aiming to improve consensus and address the limitations of existing systems. Another work focuses on enhancing movie recommender systems by incorporating visual data, specifically movie posters, to tackle the challenges of information overload, sparsity, and cold-start issues [2]. The other study introduces a recommendation system tailored for ephemeral groups attending the cinema, using the Slope One algorithm for individual predictions and the Multiplicative Utilitarian Strategy for group recommendations [3].

Some studies propose the use of ML and NLP techniques for this task. For example, a novel approach to movie recommendations [6] incorporates a knowledge graph that captures human emotions from movie reviews using machine learning techniques. By integrating users' emotional states, extracted from chat messages, with this graph, a chatbot prototype effectively tailors movie suggestions. A similar study [7] presented a model that uses social networks and microblogging data and sentiment analysis to enhance program recommendations on online media sites like YouTube and Hulu, addressing the "cold-start" problem by mining user preferences for similarity between movies and TV episodes.

Multiple studies focused on genre classification as a basis for proving movie recommendations for groups [4-5].

A hybrid approach to movie recommendations was proposed, integrating tags and human ratings to address the limitations of current services that often overlook the depth of user annotations [8]. Some other approaches include a combination of k-means clustering and genetic algorithms[9], a user-based collaborative filtering method, calculating user similarity by integrating both ratings and social connections [10], graph attention network (GAT) [11], CNN-based deep learning model that integrates basic movie attributes such as genre, cast, director, keywords, and descriptions, ratings [12].

As we see, a limited number of studies focused on consensus-oriented group recommendation of movies based on emotional features. Recognizing this issue, this paper introduces a novel methodology using emotional analysis to facilitate the group movie selection process. By analyzing emotional data from various sources, including film posters, main soundtrack, and description, our method offers a comprehensive view of the emotional landscape associated with each movie.

This approach is grounded in the understanding that films are designed to evoke specific emotional responses. Posters usually encapsulate the essence of the film, soundtracks enhance the emotional tone, and descriptions provide the context, and together, they form a multifaceted emotional profile. By integrating these elements, our methodology aims to match a film's emotional tone with the collective mood and preferences of the group, thereby simplifying the decision-making process.

The research aims to answer the following questions:
1. How do individual film preferences affect collective decision-making?
2. How do emotions from different movie sources (like posters, soundtracks, and descriptions) help us understand how a movie makes people feel?

The main contributions of this work are:
1. Proposing a novel approach for achieving consensus in group movie selection using a multi-channel emotion recognition approach based on movie data like poster, text description, and soundtrack.
2. Exploring the extent of influence various sources have on the outcome, this study investigates which element—emotions conveyed through poster colors (color image features), the soundtrack (music features), or the movie description (text features)—has the greatest impact on movie recommendations.

The paper is organized as follows. The present section is an Introduction and contains a literature review on group movie recommendations. Section II describes the methods and details of the data collection. Section III presents the experimental results and examples. Conclusion and ideas for future development are included in Section IV.

## II. METHODS

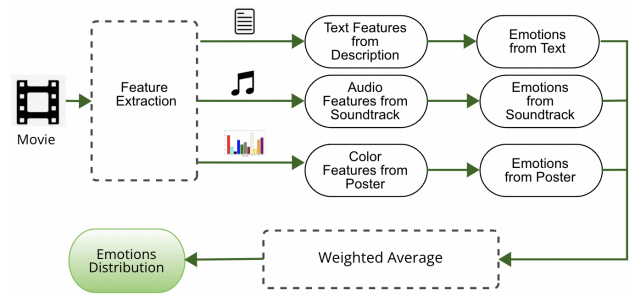

FIGURE 1. PROPOSED APPROACH FOR GROUP MOVIE RECOMMENDATIONS BASED ON EMOTION RECOGNITION

The proposed approach is presented in Figure 1. As can be seen, the emotions present in the movie are detected from three channels: Description (text features), Soundtrack (audio features), and Poster (color features). They are then aggregated using the weighted average.

### A. Data Collection

We have gathered data from the Internet Movie Database (IMDb) [15], an online repository of information about movies, television shows, and other useful information associated with them. We chose 12 movies from different genres, including comedy, romance, horror, drama, fantasy, and various years. The three main objects for each movie were collected from the database: the storyline of the movies, posters, and one of the soundtracks' excerpts with a 30-second duration. Finally, those films were selected for further analysis: Titanic (1997), Bride Wars (2009), Insidious: Chapter 3 (2015), Annabelle: Creation (2017), Just Go With It (2011), Me Before You (2016), Interstellar (2014), Edge of Tomorrow (2014), Passengers (2016), Don't Breathe 2 (2021), The Proposal (2009), and The Holiday (2006).

### B. Emotion Detection

#### a. Emotion detection in movie description

We used text2emotion 0.0.5 [13] to detect emotions presented in the movie description. Five emotion categories were used, including Happy, Angry, Sad, Surprise, and Fear. Text2emotions provides a dictionary containing keys as emotion categories and values as scores for each emotion category. Although there are five emotions, not all are represented in the text. We only consider those with a non-zero score.

#### b. Emotion detection in poster

To detect emotions from colors present in a movie poster, we employed the novel fuzzy sets-based method for categorizing emotions [14], which fits well with the imprecise and subjective nature of human assessments. It is easy to modify the suggested approach to fit our situation. The study uses fuzzy colors[18] (n=120) and range of emotions (n=10) to get fuzzy color distributions for ten emotions (anger, shyness, happiness, sadness, gratitude, shame, fear, trust, love, and surprise). Ultimately, they are transformed into a crisp domain, gaining a knowledge base of primary color-to-emotion correlations. The study identified the strong correlations between particular emotions and colors (2AFC score=0.77). We utilize these correlations and Jaccard's similarity to detect the emotions in the poster image. Considering our context, we use a subset of emotions from [10]: anger, happiness, sadness, fear, love, and surprise.

### c. Emotion detection in movie soundtrack

Low-level music features including beat, pitch, rhythm, valence, and tempo are used in the music emotion recognition [17]. Soundtrack analysis was performed using a pre-trained model based on a deep neural network algorithm, performed in [16], that obtained an F1 score of 0.91 on the authors' test set. The model is trained to recognize eight emotion categories (neutral, calm, happy, sad, angry, fearful, disgust, and surprise). We divided the initial 30-second duration audio files into ten partitions for accurate analysis. For every audio file that is supplied as input, the network may process vectors containing 40 audio features, which represent the audio frame's numerical form. We use the main five emotions (Happy, Angry, Sad, Surprise, and Fear) from [16], which is our main focus. As output, we got the emotional class suitable for the input audio excerpts, which is properly encoded (Happy=2; Sad=3; Angry=4; Fearful=5; Surprised=7). So, using ten emotional labels observed in each of the ten partitions of one whole excerpt, we formulate the dictionary with scores based on prevalence for every emotion category and keys representing emotion categories like the output of the text emotion analysis. An example is provided in Fig. 2

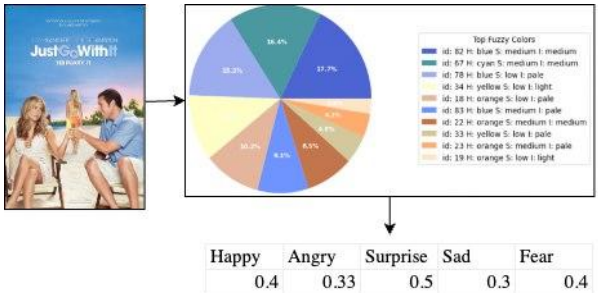

FIGURE 2. EXAMPLE OF EMOTION EXTRACTION FROM IMAGE. METHOD IS ADAPTED FROM [14]

### C. Jaccard Similarity

We calculated the Jaccard similarity coefficient, using Equation 1, between the emotional composition of each film in our database (set 1) and the best-loved film of each participant (set 2). This process gives a Jaccard value indicating the similarity between the emotional composition of the $n$ films and the participants' preferred choices. We also use it to evaluate the performance of our method by finding the similarity between real and predicted ratings.

$$J(A, B) = \frac{|A \cap B|}{|A \cup B|} \quad (1)$$

where $J$ = Jaccard distance, $A$ = set 1, $B$ = set 2. Emotion was included to the set using the following threshold for the emotion score - $th$reshold=0.1. Analyzing the distribution of the emotion scores (see Figure 3) for our predicted values, we observed that each emotion has a significance of the first quartile(Q1) of more than 0.1. So, the threshold was selected as 0.1.

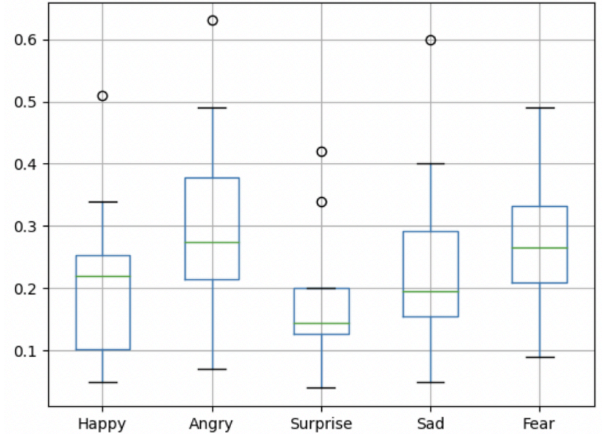

FIGURE 3. PREDICTED EMOTION SCORE DISTRIBUTION

### D. Aggregation

*1.* Aggregation of emotion scores from different channels:

The emotional scores extracted from the film poster, main soundtrack, and film description are aggregated using a weighted sum approach. Let $E_p$, $E_m$ and $E_d$ represent the emotional scores from the poster, main soundtrack, and film description, respectively. The aggregated emotional score $E_{agg}$ for each film is calculated as follows:

$$E_{agg} = \frac{w_p * E_p + w_m * E_m + w_d * E_d}{w_p + w_m + w_d} \quad (2)$$

where $w_p$, $w_m$ and $w_d$ mean the weights assigned to the emotional scores from the poster, main soundtrack, and film description, respectively. We selected the following weights based on our subjective observations: $w_p = 1$, $w_m = 2$, $w_d = 3$.

*2.* Aggregation Jaccard Similarity

The Jaccard values obtained for each participant's best-loved film are aggregated for each of the $n$ films. The aggregated Jaccard value $J^i_{agg}$ for each $i$ film is determined by averaging the Jaccard values across all participants. Let $J^i_j$ represent the Jaccard value for film $i$ obtained from participant $j$ preferences. The aggregated Jaccard value $J^i_{agg}$ for the film $i$ is calculated as follows:

$$J^i_{agg} = \frac{\sum_{j=1}^{m} J^i_j}{m} \quad (3)$$

where $m$ denotes the total number of participants, $j$ - from 1 to $m$, $i$ - from 1 to $n$.

## III. EXPERIMENTAL RESULTS
### A. Performance Evaluation

We conducted a survey to analyze participants' emotional responses to 12 films shown in Table 1, with reactions categorized as *Happy, Anger, Surprise, Sad, and Fear*. Respondents were instructed to indicate only the films that they watched and to select all applicable emotions that they experienced while watching, permitting multiple emotional responses for each question (see Fig. 4). As a result, we obtained human ratings for emotion distribution in the movie. The specific emotion score for a certain movie was calculated as the proportion of selection of this emotion among all choices made for this movie. The results are shown in Table 2.

The predicted scores for happiness, anger, surprise, sadness, and fear emotions, obtained from 3 different channels, are aggregated according to (2). The results are shown in Table 4.

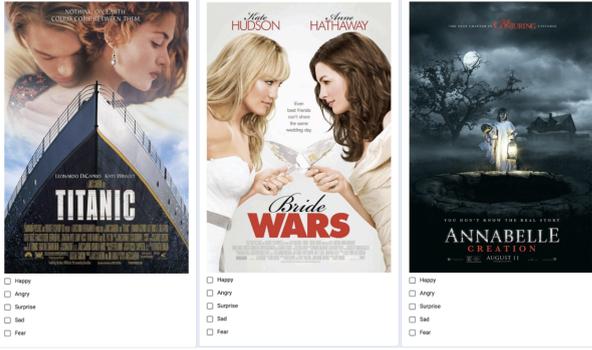

FIGURE 4. SURVEY ON MOVIE EMOTION RECOGNITION

TABLE 2. EMOTION SCORE OF SURVEY PARTICIPANTS

| id | Movie | Emotion score of survey participants |
|---|---|---|
| 1 | Titanic | {'Happy': 0.12, 'Angry': 0.07, 'Surprise': 0.1, 'Sad': 0.45, 'Fear': 0.25} |
| 2 | Bride wars | {'Happy': 0.45, 'Angry': 0.14, 'Surprise': 0.27, 'Sad': 0.05, 'Fear': 0.09} |
| … | … | … |
| 12 | The holiday | {'Happy': 0.55, 'Angry': 0.05, 'Surprise': 0.18, 'Sad': 0.14, 'Fear': 0.09} |

TABLE 1. INPUT PARAMETERS

| id | Movie | Description text | Soundtrack | Poster |
|---|---|---|---|---|
| 1 | Titanic | 84 years later, a 100 year-old woman named Rose DeWitt Bukater tells the story to her granddaughter Lizzy Calvert, … | My Heart Will Go On by Celine Dion | |
| 2 | Bride wars | In Manhattan, the lawyer Liv and the school teacher Emma have been best friends since their childhood. They both are proposed … | Somethin' Special by Colbie Caillat | |
| … | … | … | … | … |
| 12 | The holiday | Iris is in love with a man who is about to marry another woman. Across the globe, Amanda realizes the man she lives with has been unfaithful... | Last Christmas by Wham! | |

Now, we can compare the prediction power of our method by finding the similarity index between predicted emotion scores and the scores given by survey participants. Using the aggregated results of predicted scores (see Table 4) and real human scores (Table 2), we calculated the Jaccard distance (1) between the real and the expected values of emotions for each movie (see Table 3). Afterward, we estimated the average similarity coefficient to be 0.58.

### B. Example: Group Movie Selection

Let us show the example of selecting a movie for a group of people using the proposed approach. Given four movie viewers and a pool of 12 movie options (shown in Table 1), we aim to ensure satisfaction and provide the best movie recommendation for the participants.

Each participant provided information about their best-loved film, a reference point for their preferences (see Table 5). Using the methods described in Section II, we conducted an emotional analysis of three primary sources associated with each film: the film poster (picture), the main soundtrack (music), and the film description (text). Emotional data containing - happiness, anger, surprise, sadness, and fear from the input are shown in Table 4, and emotional data of best-loved films of each participant is provided in Table 5.

We calculated the Pearson correlation coefficient between each emotion channel (text, colors, audio) and real human ratings to evaluate which emotion channel has the biggest impact on human impression. The highest correlation is between human ratings and text description emotion channel (0.43).

TABLE 3. JACCARD SIMILARITY INDEX BETWEEN PREDICTED AND REAL EMOTION DISTRIBUTION FOR EACH MOVIE. THE MEAN JACCARD SIMILARITY INDEX IS 0.58.

| Movie | Titanic | Bride wars | Insidious 3 | Annabelle: Creation | Just go with it | Me before you | Interstellar | Edge of tomorrow | Passengers | Don't breathe 2 | The Proposal | The holiday |
|---|---|---|---|---|---|---|---|---|---|---|---|---|
| Jaccard Similarity | 0.6 | 0.6 | 0.25 | 0.4 | 0.5 | 0.6 | 1.0 | 1.0 | 0.4 | 0.5 | 0.5 | 0.6 |

TABLE 4. PREDICTED EMOTION SCORE OF FILMS

| id | Movie | Emotion score of films from description text | Emotion score of films from soundtrack | Emotion score of films from poster | Average emotion score of films |
|---|---|---|---|---|---|
| 1 | Titanic | {'Happy': 0.17, 'Angry': 0.0, 'Surprise': 0.0, 'Sad': 0.33, 'Fear': 0.5} | {'Happy': 0.0, 'Angry': 0.33, 'Surprise': 0.0, 'Sad': 0.33, 'Fear': 0.33} | {'Happy': 0.78, 'Angry': 0.56, 'Surprise': 0.75, 'Sad': 0.67, 'Fear': 0.78} | {'Happy': 0.22, 'Angry': 0.2, 'Surprise': 0.13, 'Sad': 0.39, 'Fear': 0.49} |
| 2 | Bride wars | {'Happy': 0.75, 'Angry': 0.0, 'Surprise': 0.0, 'Sad': 0.0, 'Fear': 0.25} | {'Happy': 0.13, 'Angry': 0.63, 'Surprise': 0.0, 'Sad': 0.25, 'Fear': 0.0} | {'Happy': 0.56, 'Angry': 0.33, 'Surprise': 0.71, 'Sad': 0.44, 'Fear': 0.56} | {'Happy': 0.51, 'Angry': 0.27, 'Surprise': 0.12, 'Sad': 0.16, 'Fear': 0.22} |
| … | … | … | … | … | … |
| 12 | The holiday | {'Happy': 0.38, 'Angry': 0.08, 'Surprise': 0.08, 'Sad': 0.31, 'Fear': 0.15} | {'Happy': 0.0, 'Angry': 0.0, 'Surprise': 0.0, 'Sad': 1.0, 'Fear': 0.0} | {'Happy': 0.78, 'Angry': 0.75, 'Surprise': 0.56, 'Sad': 0.67, 'Fear': 0.78} | {'Happy': 0.32, 'Angry': 0.17, 'Surprise': 0.13, 'Sad': 0.6, 'Fear': 0.21} |

TABLE 5. EMOTION SCORE OF PARTICIPANTS' FAVORITE FILMS

| id | Movie | Emotion score of films from description text | Emotion score of films from soundtrack | Emotion score of films from poster | Average emotion score of films |
|---|---|---|---|---|---|
| 1 | The Notebook | {'Happy': 0.45, 'Angry': 0.05, 'Surprise': 0.15, 'Sad': 0.25, 'Fear': 0.1} | {'Happy': 0.25, 'Angry': 0.0, 'Surprise': 0, 'Sad': 0.5, 'Fear': 0.25} | {'Happy': 0.56, 'Angry': 0.33, 'Surprise': 0.71, 'Sad': 0.44, 'Fear': 0.56} | {'Happy': 0.4, 'Angry': 0.08, 'Surprise': 0.19, 'Sad': 0.37, 'Fear': 0.23} |
| 2 | Split | {'Happy': 0.0, 'Angry': 0.22, 'Surprise': 0.11, 'Sad': 0.22, 'Fear': 0.44} | {'Happy': 0.5, 'Angry': 0.25, 'Surprise': 0, 'Sad': 0.25, 'Fear': 0.0} | {'Happy': 0.56, 'Angry': 0.5, 'Surprise': 0.33, 'Sad': 0.44, 'Fear': 0.56} | {'Happy': 0.26, 'Angry': 0.28, 'Surprise': 0.11, 'Sad': 0.27, 'Fear': 0.31} |
| 3 | Oppenheimer | {'Happy': 0.25, 'Angry': 0.0, 'Surprise': 0.25, 'Sad': 0.0, 'Fear': 0.5} | {'Happy': 0.0, 'Angry': 1.0, 'Surprise': 0, 'Sad': 0.0, 'Fear': 0.0} | {'Happy': 0.33, 'Angry': 0.43, 'Surprise': 0.25, 'Sad': 0.38, 'Fear': 0.33} | {'Happy': 0.18, 'Angry': 0.41, 'Surprise': 0.17, 'Sad': 0.06, 'Fear': 0.31} |
| 4 | Barbie | {'Happy': 0.06, 'Angry': 0.03, 'Surprise': 0.09, 'Sad': 0.41, 'Fear': 0.41} | {'Happy': 0.0, 'Angry': 0.0, 'Surprise': 1, 'Sad': 0.0, 'Fear': 0.0} | {'Happy': 0.36, 'Angry': 0.3, 'Surprise': 0.3, 'Sad': 0.27, 'Fear': 0.36} | {'Happy': 0.09, 'Angry': 0.07, 'Surprise': 0.43, 'Sad': 0.25, 'Fear': 0.27} |

Emotional scores from the three sources were aggregated using Eq. 2, for each of the 12 offered movies and four favorite movies. After that, we calculated the Jaccard similarity coefficient between each film's emotional composition and each participant's best-loved film. This process gives a Jaccard value indicating the similarity between the emotional composition of the 12 films and the participants' preferred choices, shown in Tables 4 and 5. The Box plot of predicted emotion scores is shown in Fig. 3.

TABLE 6. INPUT PARAMETERS OF PARTICIPANTS' FAVORITE FILMS

| id | Movie | Description text | Soundtrack | Poster |
|---|---|---|---|---|
| 1 | The Notebook | With almost religious devotion, Duke, a kind octogenarian inmate of a peaceful nursing home, reads daily a captivating story… | I'll Be Seeing You by Billie Holiday | 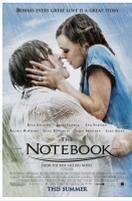 |
| 2 | Split | Though Kevin (James McAvoy) has evidenced 23 personalities to his trusted psychiatrist, Dr. Fletcher (Betty Buckley), there … | In September by Slam Allen | 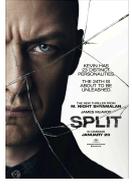 |
| 3 | Oppenheimer | A dramatization of the life story of J. Robert Oppenheimer, the physicist who had a large hand in the development of … | Can You Hear The Music by Ludwig Göransson | 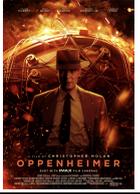 |
| 4 | Barbie | Barbie the Doll lives in bliss in the matriarchal society of Barbieland feeling good about her role in the world in the various iterations of Barbies … | Dance the Night by Dua Lipa | 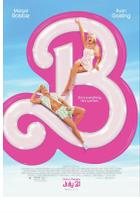 |

The set of films with the mean Jaccard value equal to the highest coefficient, having the greatest similarity to the participants' best-loved films, was identified as the best choice for the group. After that, we filtered them by genre. As a result, we obtained two movies with the highest recommendation scores of 0.8, namely "Titanic" and "Me Before You." The least recommended movie for this group is "Passengers," with a score of 0.34.

## IV. Conclusion

In this paper, we provide a multi-modal emotion recognition approach that uses emotional analysis to identify the most suitable film for a group of participants. Our method integrates emotional data from multiple sources, including film posters, main soundtracks of film, and film descriptions, to provide a comprehensive understanding of the emotional landscape associated with each film.

Emotional analysis represents a promising approach to facilitate consensus and satisfaction in group film selection scenarios. By incorporating emotional data from multiple channels and employing the Jaccard similarity coefficient, our methodological approach provides a method for optimal film selection that considers individual preferences and emotional components of films.

In future works, we plan to collect bigger datasets and conduct more extensive experiments with real groups of people. In addition, we plan to integrate the genre information into the algorithm.

## References

[1] L. Quijano-Sanchez, J. A. Recio-Garcia and B. Diaz-Agudo, "HappyMovie: A Facebook Application for Recommending Movies to Groups," 2011 IEEE 23rd International Conference on Tools with Artificial Intelligence, Boca Raton, FL, USA, 2011, pp. 239-244, doi: 10.1109/ICTAI.2011.44.

[2] A. F. Rahmatabadi, A. Bastanfard, A. Amini and H. Saboohi, "Building Movie Recommender Systems Utilizing Poster's Visual Features: A Survey Study," 2022 10th RSI International Conference on Robotics and Mechatronics (ICRoM), Tehran, Iran, Islamic Republic of, 2022, pp. 110-116, doi: 10.1109/ICRoM57054.2022.10025210.

[3] G. Fernández, W. López, F. Olivera, B. Rienzi and P. Rodríguez-Bocca, "Let's go to the cinema! A movie recommender system for ephemeral groups of users," 2014 XL Latin American Computing Conference (CLEI), Montevideo, Uruguay, 2014, pp. 1-12, doi: 10.1109/CLEI.2014.6965161.

[4] W. Li, J. Xu, Q. Bao, R. Shen, H. Yuan and M. Xu, "An Adaptive Aggregation Method Based on Movie Genre for Group Recommendation," 2020 IEEE 32nd International Conference on Tools with Artificial Intelligence (ICTAI), Baltimore, MD, USA, 2020, pp. 73-78, doi: 10.1109/ICTAI50040.2020.00022.

[5] Sang-Min Choi, Sang-Ki Ko, Yo-Sub Han, A movie recommendation algorithm based on genre correlations, Expert Systems with Applications, Volume 39, Issue 9, 2012, Pages 8079-8085, https://doi.org/10.1016/j.eswa.2012.01.132.

[6] Arno Breitfuss, Karen Errou, Anelia Kurteva, Anna Fensel, Representing emotions with knowledge graphs for movie recommendations, Future Generation Computer Systems, Volume 125, 2021, Pages 715-725, https://doi.org/10.1016/j.future.2021.06.001.

[7] Hui Li, Jiangtao Cui, Bingqing Shen, Jianfeng Ma, An intelligent movie recommendation system through group-level sentiment analysis in microblogs, Neurocomputing, Volume 210, 2016, Pages 164-173, https://doi.org/10.1016/j.neucom.2015.09.134.

[8] Shouxian Wei, Xiaolin Zheng, Deren Chen, Chaochao Chen, A hybrid approach for movie recommendation via tags and ratings, Electronic Commerce Research and Applications, Volume 18, 2016, Pages 83-94, https://doi.org/10.1016/j.elerap.2016.01.003.

[9] Zan Wang, Xue Yu, Nan Feng, Zhenhua Wang, An improved collaborative movie recommendation system using computational intelligence, Journal of Visual Languages & Computing, Volume 25, Issue 6, 2014, Pages 667-675, https://doi.org/10.1016/j.jvlc.2014.09.011.

[10] Aamir Fareed, Saima Hassan, Samir Brahim Belhaouari, Zahid Halim, A collaborative filtering recommendation framework utilizing social networks, Machine Learning with Applications, Volume 14, 2023, 100495, https://doi.org/10.1016/j.mlwa.2023.100495.

[11] W. -H. Liao, Y. -T. Lin, C. -Y. Lin and S. -C. Kuai, "A Group Recommendation System for Movies Using Deep Learning," 2023 International Conference on Consumer Electronics - Taiwan (ICCE-Taiwan), PingTung, Taiwan, 2023, pp. 61-62, doi: 10.1109/ICCE-Taiwan58799.2023.10226648.

[12] S. Sahu, R. Kumar, M. S. Pathan, J. Shafi, Y. Kumar and M. F. Ijaz, "Movie Popularity and Target Audience Prediction Using the Content-Based Recommender System," in IEEE Access, vol. 10, pp. 42044-42060, 2022, doi: 10.1109/ACCESS.2022.3168161.

[13] LLC, M. (2020). text2emotion 0.0.5 package.

[14] Muratbekova, M., & Shamoi, P. (2024). Color-Emotion Associations in Art: Fuzzy Approach. IEEE ACCESS.

[15] Internet Movie Database (IMDb), https://www.imdb.com.

[16] M. G. de Pinto, M. Polignano, P. Lops and G. Semeraro, "Emotions Understanding Model from Spoken Language using Deep Neural Networks and Mel-Frequency Cepstral Coefficients," 2020 IEEE Conference on Evolving and Adaptive Intelligent Systems (EAIS), Bari, Italy, 2020, pp. 1-5, doi: 10.1109/EAIS48028.2020.9122698.

[17] A. Ualibekova and P. Shamoi, "Music Emotion Recognition Using K-Nearest Neighbors Algorithm," 2022 International Conference on Smart Information Systems and Technologies (SIST), Nur-Sultan, Kazakhstan, 2022, pp. 1-6, doi: 10.1109/SIST54437.2022.9945814.

[18] P. Shamoi, D. Sansyzbayev and N. Abiley, "Comparative Overview of Color Models for Content-Based Image Retrieval," 2022 International Conference on Smart Information Systems and Technologies (SIST), Nur-Sultan, Kazakhstan, 2022, pp. 1-6, doi: 10.1109/SIST54437.2022.9945709.